\documentclass[compsoc,conference,a4paper,10pt,times]{IEEEtran}
\usepackage[backend=bibtex,
    bibstyle=ieee,
    sortcites=true,
    maxnames=2,
    maxbibnames=12,
    minnames=1,
    defernumbers=true,
    uniquelist=false,
]{biblatex}

\usepackage{ifthen}
\makeatletter
\newcounter{IEEE@bibentries}
\renewcommand\IEEEtriggeratref[1]{%
  \renewbibmacro{finentry}{%
    \stepcounter{IEEE@bibentries}%
    \ifthenelse{\equal{\value{IEEE@bibentries}}{#1}}
    {\finentry\@IEEEtriggercmd}
    {\finentry}%
  }%
}
\makeatother

\usepackage[colorlinks=true,urlcolor=black]{hyperref}

\def\BibTeX{{\rm B\kern-.05em{\sc i\kern-.025em b}\kern-.08em
    T\kern-.1667em\lower.7ex\hbox{E}\kern-.125emX}}

\usepackage{booktabs}
\usepackage{tabularx}
\usepackage{csquotes}
\usepackage{tikz}
\usepackage{pgfplots}
\usetikzlibrary{
        shapes,
        decorations
    }

\usepackage{amsmath,amssymb,amsfonts}
\usepackage{cleveref}
\crefname{figure}{\figurename}{Figures}
\crefname{table}{\tablename}{Tables}
\usepackage{algorithmic}
\usepackage{graphicx}
\usepackage{textcomp}
\usepackage{xcolor}
\usepackage{xspace}
\newcommand*{\eg}{e.g.\@\xspace}
\newcommand*{\ie}{i.e.\@\xspace}

\definecolor{sron0}{HTML}{332288}
\definecolor{sron1}{HTML}{88CCEE}
\definecolor{sron2}{HTML}{117733}
\definecolor{sron3}{HTML}{DDCC77}
\definecolor{sron4}{HTML}{CC6677}
\definecolor{sron5}{HTML}{AA4499}

\begin{document}

\title{Towards Verifiable Mutability for Blockchains}

\author{\IEEEauthorblockN{Erik Daniel}
\IEEEauthorblockA{\textit{Distributed Security Infrastructures} \\
\textit{Technische Universität Berlin}\\
 % Berlin, Germany \\
erik.daniel@tu-berlin.de}
\and
\IEEEauthorblockN{Florian Tschorsch}
\IEEEauthorblockA{\textit{Distributed Security Infrastructures} \\
\textit{Technische Universität Berlin}\\
 % Berlin, Germany \\
florian.tschorsch@tu-berlin.de}
}

\maketitle

\begin{abstract}
Due to their immutable log of information,
blockchains can be considered as a transparency-enhancing technology.
The immutability, however, also introduces threats and challenges
with respect to privacy laws and illegal content.
Introducing a certain degree of mutability,
which enables the possibility to store and remove information,
can therefore increase the opportunities for blockchains.
In this paper, we present a concept for a mutable blockchain structure.
Our approach enables the removal of certain blocks,
while maintaining the blockchain's verifiability property.
Since our concept is agnostic to any consensus algorithms,
it can be implemented with permissioned and permissionless blockchains.
\end{abstract}

\section{Introduction}\label{sec:intro}
Blockchains provide many useful properties,
ideally reducing trust assumptions.
The possibility to retrieve data and verify its validity locally
makes blockchains a transparency-enhancing technology.

While blockchains offer transparency,
the immutability property brings its own opportunities,
challenges and threats~\cite{casino2019immutability,matzutt2018quantitative}.
One big challenge is the inclusion of personal data.
The EU General Data Protection Regulation (GDPR)
allows a data subject the \enquote{right to erasure}
as well as the \enquote{right to rectification} without undue delay.
Therefore, the immutability prevents or at least hinders
the usage of blockchains in cases where personal information is involved.
Some solutions propose to store on-chain storage locations and hashes
of data, storing the real data off-chain.
In these cases, the blockchchain provides verifiability for the data.
However, the hash pointers can still be classified as personal data, \ie, a pseudonym.

In this concept paper, we propose a blockchain-like structure that enables
the attachment and removal of information.
To this end, we introduce intentional break points in the blockchain
that allow us to remove certain blocks,
so-called removable blocks.
The design combines aspects of skipchains~\cite{nikitin2017chainiac}
as well as redactable blockchains~\cite{deuber2019redactable}.
In our approach, however, the delete operation is initiated through a transaction,
making the deletion process verifiable.
While our approach removes the immutability of some blocks,
we maintain chain integrity and verifiability.
Since our approach builds upon a general blockchain structure,
it can extend existing designs, \eg Bitcoin~\cite{nakamoto2008bitcoin}.

For a better understanding of the concept,
we apply it to the use case of consent management.
Moreover, we qualitatively discuss concerns and risks that arise from
the mutability and specifically the deletion process.

The main contributions of this paper are
(i)~a structure for a mutable blockchain that maintains the chain's verifiability (\Cref{sec:concept}),
(ii)~a mechanism to store data on-chain in a removable way (\Cref{sec:delete}),
(iii)~a use case for the structure in the form of explicit blockchain-based consent management (\Cref{sec:consent}),
and (iv)~a discussion of risks (\Cref{sec:analysis}).

\section{Related Work}\label{sec:relwork}
Relaxing the immutability of a blockchain is sometimes a desirable property.
While an immutable blockchain maintains integrity,
it prevents removing illegal or malicious content,
which should not be replicated~\cite{matzutt2018quantitative}.
Removing the immutability property, however, also introduces new problems.
One problem is breaking the chain consistency by removing a block.
Every block typically points to its predecessor by referencing its hash.
Removing a block breaks the link, rendering it impossible to follow the chain from start to end.
\citeauthor{deuber2019redactable}~\cite{deuber2019redactable} address this problem
with a multi-link blockchain using a voting mechanism to replace blocks, which breaks one of the links.
\citeauthor{ateniese2017redactable}~\cite{ateniese2017redactable} use chameleon hashes,
allowing the computation of a hash collision through a trapdoor,
making restoration of the link possible.
Recent research~\cite{dousti2020moderated} indicates
possible problems with these forms of redactable blockchains.

Another problem concerns the state of the blockchain resulting from block removal.
Blocks typically hold multiple transactions,
which are potentially all affected by a block removal.
For example, in a rolling blockchain~\cite{dennis2016temporal},
older blocks are removed to maintain a fixed number of blocks in the chain.
Due to the removal of old blocks, transactions reference unknown inputs and
it becomes impossible to verify the validity of the transaction.
Another example are hard forks, where the blockchain starting from a specific block
is cut and transactions thereby reversed.
Hard forks are mainly used for substantial protocol changes, but they can also be used
to undo malicious behavior, \eg, as in the case of the Ethereum DAO
hack and its resulting hard fork~\cite{mehar2019understanding}.
Especially in case of hard forks the disappearance of transactions
is a minor problem, since all following transactions are removed
and the verifiability of previous transactions is secured.
Pruning is another method for removing transactions.
Although, pruning is mainly used for saving storage by deleting old transactions
from blocks~\cite{florian2019erasing},
and like with a rolling chain transaction inputs become unknown.
Nevertheless, removing transactions can lead to undesired problems, as shown by
\citeauthor{puddu2017muchain}~\cite{puddu2017muchain} in $\mu$chain,
in which transactions can be modified after their verification.
A more detailed overview of blockchain mutability
challenges and possible solutions can be found in~\cite{politou2019blockchain}.

The data structure of our proposal is similar to a skipchain~\cite{nikitin2017chainiac},
allowing to delete certain blocks while maintaining a chain.
The idea is also similar to the redactable blockchain
with multiple links~\cite{deuber2019redactable}.
Instead of replacing blocks and voting for the acceptance, however,
we distinguish between removable and permanent blocks,
and blocks are removed with transactions included in permanent blocks
providing confirmation of missing blocks.

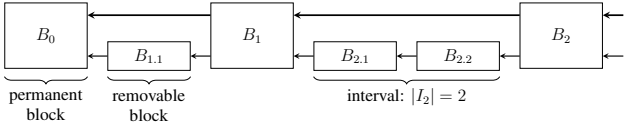
\begin{figure}[t]
\begin{center}
  \resizebox{\columnwidth}{!}{
  \begin{tikzpicture}[every node/.style={font=\large},node distance=25mm]
  \tikzstyle{pblock} = [rectangle, minimum width=20mm, minimum height=16mm,text centered, draw=black, align=left]
  \tikzstyle{dblock} = [rectangle, minimum width=20mm, minimum height=7mm,text centered, draw=black, align=left]

      \node (b0) [pblock] {$B_{0}$};
      \node (b1) [dblock, right of = b0,yshift=-5mm] {$B_{1.1}$};
      \node (b3) [pblock, right of = b1,yshift=5mm] {$B_{1}$};
      \node (b4) [dblock, right of = b3,yshift=-5mm] {$B_{2.1}$};
      \node (b5) [dblock, right of = b4] {$B_{2.2}$};
      \node (b6) [pblock, right of = b5,yshift=5mm] {$B_{2}$};

      \draw [thick,->,>=stealth] (b1.west) -- ([yshift=-5mm]b0.east);
      \draw [very thick,->,>=stealth] ([yshift=5mm]b3.west) -- ([yshift=5mm]b0.east);
      \draw [thick,->,>=stealth] ([yshift=-5mm]b3.west) -- (b1.east);
      \draw [thick,->,>=stealth] (b4.west) -- ([yshift=-5mm]b3.east);
      \draw [thick,->,>=stealth] (b5.west) -- (b4.east);
      \draw [very thick,->,>=stealth] ([yshift=5mm]b6.west) -- ([yshift=5mm]b3.east);
      \draw [thick,->,>=stealth] ([yshift=-5mm]b6.west) -- (b5.east);
      \draw [very thick,<-,>=stealth] ([yshift=5mm]b6.east) -- ([yshift=5mm,xshift=5mm]b6.east);
      \draw [thick,<-,>=stealth] ([yshift=-5mm]b6.east) -- ([yshift=-5mm,xshift=5mm]b6.east);

  \draw [decorate,decoration={brace,amplitude=5pt,raise=20pt,mirror}]
    ([yshift=-3mm]b0.west) -- ([yshift=-3mm]b0.east)
    node [black,midway,yshift=-14mm,text width=20mm,text centered] {permanent block};

  \draw [decorate,decoration={brace,amplitude=5pt,raise=20pt,mirror}]
    ([yshift=2mm]b1.west) -- ([yshift=2mm]b1.east)
    node [black,midway,yshift=-14mm,text width=20mm,text centered] {removable block};

  \draw [decorate,decoration={brace,amplitude=5pt,raise=20pt,mirror}]
    ([yshift=2mm]b4.west) -- ([yshift=2mm]b5.east)
    node [black,midway,yshift=-12mm] {interval: $|I_{2}| = 2$};
  \end{tikzpicture}}
\end{center}
\vspace{-1em}
\caption{Visualization of the blockchain structure with a varying interval.}
\label{fig:chainexample}
\end{figure}

\section{Mutable Blockhain Structure}\label{sec:concept}
In this section, we give a general overview of our concept for a mutable blockchain-based structure,
enabling removal of certain blocks without destroying the chain's verifiability.
Removing blocks allows to \enquote{forget} data.
Transactions initiate the removal of specific blocks.

\subsection{Data Structure}
In our design, we differentiate between two types of blocks,
\emph{permanent blocks} and \emph{removable blocks},
which is illustrated in \cref{fig:chainexample}.
Permanent blocks~$B_{i}$ use a hash pointer to reference a permanent block \emph{and} a removable block.
Removable blocks~$B_{i.j}$, on the other hand, refer only to one previous block.
The structure is similar to the data structure of a skip list,
and related to~\cite{nikitin2017chainiac,deuber2019redactable}.
The number of removable blocks between two consecutive permanent blocks is called \emph{interval}.
The interval $I_{i}$ denotes the removable blocks~$B_{i.*}$ preceding block~$B_{i}$,
\eg, $I_{2}=\{B_{2.1},B_{2.2}\}$.
Accordingly, the interval length~$\mid I_{i} \mid$ determines the number of removable blocks.
An interval length of zero results in a blockchain consisting of permanent blocks only,
conceptually comparable to immutable blockchains.
The interval length is an adjustable parameter
included in the subsequent permanent block's header.

\begin{table}
\scriptsize
\centering
\caption{New transaction types for the structure and use case}
\begin{tabular}{llll}
\toprule
Transaction & Notation & Semantics & Block type \\
\midrule
Register & $Reg(x)$ & register entity x with $pk_{x}$ & permanent \\
Prepare & $Prep(x)$ & prepare delete of $I_{x}$ & permanent \\
Delete & $Del(x)$ & $I_{x}$ can be deleted & permanent \\
Removable & $Rem(x)$ & removable information x & removable \\
\midrule
Information & $Info(x)$ & x's consent collection purpose & permanent \\
Consent & $Con(x,y)$ & x gives consent with value x & permanent \\
\bottomrule
\end{tabular}
\label{tab:txoverview}
\end{table}

\subsection{Removable Information}
While permanent blocks contain immutably stored information,
removable blocks contain information, which should be allowed to be removed.
For our concept and our use case (see \Cref{sec:consent}), we introduce new transaction types,
which are summarized in \cref{tab:txoverview}.

As an example, assume Alice wants to provide removable information.
She therefore needs to register first with a register transaction.
We denote Alice's public key with $pk_{A}$ and
her signed transaction as $sk_{A}(Tx)$.
Accordingly, $sk_{A}(Reg(A))$ is
the register transaction for and signed by Alice.
Note, a register transaction has no input,
but requires an output that can be referenced multiple times\footnote{
Maintaining the notion of a UTXO,
a register transaction could prepare multiple outputs.
restricting the number of subsequent transactions.}.

After the registration, Alice can issue removable transactions to provide removable information.
Removable transactions are self-contained statements, \ie, independent of other transactions.
They however require the output of the register transaction as an input to create a traceable link.
Removable transactions have no output.
In general, removable transactions can contain any information
which might be sensitive if included immutably in the chain,
\eg, files, hash pointers to off-chain storage locations or
personal information about a data subject.

For example, Alice's removable transaction contains her e-mail addresses $m$,
\ie, $sk_{A}(Rem(m))$.
The removable transaction is confirmed by a removable block,
since it contains data, which should be rectifiable, erasable, or simply removable.
If Alice wants to erase her e-mail address,
the removable block containing her transaction has to (and can) be removed.

\begin{figure*}[t]
\footnotesize
\textbf{State 1:} Confirm prepare transaction for delete of $I_{1}$ \\

\begin{tikzpicture}[every node/.style={font=\scriptsize},node distance=25mm]
\tikzstyle{pblock} = [rectangle, minimum width=19mm, minimum height=22mm,text centered, draw=black, align=left]
\tikzstyle{dblock} = [rectangle, minimum width=19mm, minimum height=13mm,text centered, draw=black, align=left]
\tikzstyle{iblock} = [rectangle, minimum width=19mm, minimum height=13mm,text centered, draw=gray, dashed, align=left]

    \node (b0) [pblock] {$B_{0}$ \\ \\ $\mid I_{0} \mid =0$ \\ $P_{0}=\emptyset$ \\ \\
    $sk_{A}(Reg(A))$ \\ $sk_{B}(Reg(B))$};
    \node (b1) [dblock, right of = b0,yshift=-5mm] {$B_{1.1}$ \\ \\ $sk_{A}(Rem(m))$ \\ \textcolor{red}{$\mathbf{sk_{B}(Rem(n))}$}};
    \node (b3) [pblock, right of = b1,yshift=5mm] {$B_{1}$ \\ \\
     \\ $\mid I_{1} \mid = 1$ \\ $P_{1}=\{pk_{A},$ \\ $pk_{B}\}$ \\ \\};
    \node (b4) [dblock, right of = b3,yshift=-5mm, very thick, dotted] {$B_{2.1}$ \\ \\ \textcolor{red}{$\mathbf{sk_{B}(Rem(n))}$}};
    \node (b6) [pblock, right of = b4,yshift=5mm, very thick,dotted] {$B_{2}$ \\ \\ $\mid I_{2} \mid =1$ \\ $P_{2}=\{pk_{B}\}$ \\ \\ \textcolor{red}{$\mathbf{sk_{A}(Prep(1))}$}};

    \draw [thick,->,>=stealth] (b1.west) -- ([yshift=-5mm]b0.east);
    \draw [very thick,->,>=stealth] ([yshift=5mm]b3.west) -- ([yshift=5mm]b0.east);
    \draw [thick,->,>=stealth] ([yshift=-5mm]b3.west) -- (b1.east);
    \draw [thick,->,>=stealth] (b4.west) -- ([yshift=-5mm]b3.east);
    \draw [very thick,->,>=stealth] ([yshift=5mm]b6.west) -- ([yshift=5mm]b3.east);
    \draw [thick,->,>=stealth] ([yshift=-5mm]b6.west) -- (b4.east);
\end{tikzpicture}
\vspace{0.9em}

\textbf{State 2:} Delete $I_{1}$ after additional confirmations of the delete transaction \\

\begin{tikzpicture}[every node/.style={font=\scriptsize},node distance=25mm]
\tikzstyle{pblock} = [rectangle, minimum width=19mm, minimum height=22mm,text centered, draw=black, align=left]
\tikzstyle{dblock} = [rectangle, minimum width=19mm, minimum height=13mm,text centered, draw=black, align=left]
\tikzstyle{iblock} = [rectangle, minimum width=19mm, minimum height=13mm,text centered, draw=gray, dashed, align=left]

    \node (b0) [pblock] {$B_{0}$ \\ \\ $\mid I_{0} \mid =0$ \\ $P_{0}=\emptyset$ \\ \\
    $sk_{A}(Reg(A))$ \\ $sk_{B}(Reg(B))$};
    %\node (b1) [iblock, right of = b0,yshift=-5mm] {$B_{0.1}$ \\ \\ $sk_{A}(Tx_{per}(m))$ \\ $sk_{B}(Tx_{per}(n))$};
    \node (b1) [iblock, right of = b0,yshift=-5mm,draw=none] {};
    \node (b3) [pblock, right of = b1,yshift=5mm] {$B_{1}$ \\ \\
     \\ $\mid I_{1} \mid = 1$ \\ $P_{1}=\{pk_{A},$ \\ $pk_{B}\}$};
    \node (b4) [dblock, right of = b3,yshift=-5mm] {$B_{2.1}$ \\ \\ $sk_{B}(Rem(n))$};
    \node (b6) [pblock, right of = b4,yshift=5mm] {$B_{2}$ \\ \\ $\mid I_{2} \mid =1$ \\ $P_{2}=\{pk_{B}\}$ \\ \\ $sk_{A}(Prep(1))$};
    \node (b7) [dblock, right of = b6,yshift=-5mm] {$B_{3.1}$ \\ \\ $sk_{A}(Rem(o))$};
    \node (b8) [pblock, right of = b7,yshift=5mm] {$B_{3}$ \\ \\ $\mid I_{3} \mid = 1$ \\ $P_{3}=\{pk_{A}\}$ \\ \\ \textcolor{red}{$\mathbf{sk_{A}(Del(1))}$}};

    \draw [very thick,->,>=stealth] ([yshift=5mm]b3.west) -- ([yshift=5mm]b0.east);
    \draw [thick,->,>=stealth] ([yshift=-5mm]b3.west) -- (b1.east) node [midway, cross out, very thick, draw, solid, red, inner sep=2pt] {};

    \draw [thick,->,>=stealth] (b4.west) -- ([yshift=-5mm]b3.east);
    \draw [very thick,->,>=stealth] ([yshift=5mm]b6.west) -- ([yshift=5mm]b3.east);
    \draw [thick,->,>=stealth] ([yshift=-5mm]b6.west) -- (b4.east);

    \draw [thick,->,>=stealth] (b7.west) -- ([yshift=-5mm]b6.east);
    \draw [very thick,->,>=stealth] ([yshift=5mm]b8.west) -- ([yshift=5mm]b6.east);
    \draw [thick,->,>=stealth] ([yshift=-5mm]b8.west) -- (b7.east);

    \draw [very thick,<-,>=stealth] ([yshift=5mm]b8.east) -- ([yshift=5mm,xshift=5mm]b8.east);
    \draw [thick,<-,>=stealth] ([yshift=-5mm]b8.east) -- ([yshift=-5mm,xshift=2mm]b8.east);
\end{tikzpicture}
\caption{Step-by-step example for deleting an interval.}
\vspace{-0.8em}
\label{fig:chaindelexample}
\end{figure*}
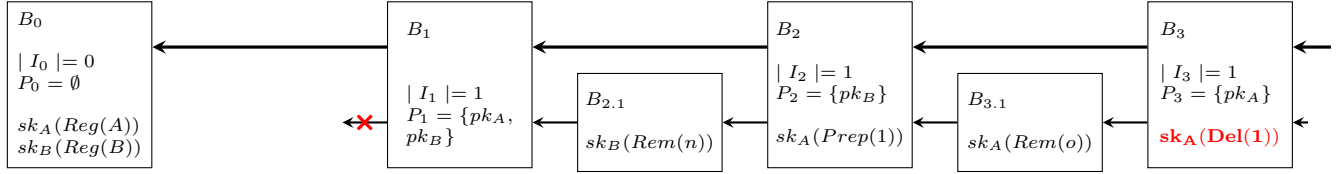

\section{Block Removal}\label{sec:delete}
In our concept, removable blocks and therefore removable transactions
can be removed by dropping complete intervals containing the respective blocks,
\eg, $B_{2.1}$ and $B_{2.2}$ for interval~$I_{2}$ in \cref{fig:chainexample}.
The removal of an interval is initiated with a \emph{delete transaction},
referencing the interval to be deleted.
Delete transactions are confirmed in permanent blocks,
allowing the verification of missing intervals.
In the following, we discuss design considerations,
before describing the removal process in detail.

\subsection{Methods}
We differentiate between authorized removal and unauthorized removal.
The method should be specified in advance,
since they enforce different block and transaction validation.
In case of an unauthorized removal, intervals can be removed
by anyone by issuing a delete transaction.
The unauthorized delete might be useful
for removing sensitive or unwanted content in an interval.
For unauthorized removal,
the miners become the judges of the validity of deletes.
A delete transaction would only be confirmed and therefore included in a block,
if they agree to the deletion.
Otherwise, the delete transaction and blocks containing the delete transaction will be ignored.

When removable transactions contain personal information,
authorized removal is preferred.
Therefore, a delete transaction needs to be signed by the issuer of the removal.
A permanent block then requires additionally a list of public keys~$P_{i}$
of all public keys of the involved removable transactions of the respective interval.
Keys in~$P_{i}$ are eligible to issue the removal of blocks in the previous interval.
Accordingly, if the key is included, $I_{i}$ includes a removable transaction of the respective issuer.

A na\"ive approach to execute a removal would be
to drop the complete interval without any precautions.
This approach is vulnerable to malicious removals, though:
An adversary could issue many removable transactions
to be allowed to remove respective intervals.
The adversary can issue removal of intervals
containing removable transactions of other entities,
effectively removing benign entities' transactions.
While the na\"ive approach preserves the privacy of the removal initiator,
\eg, by using ring signatures to authorize delete transactions,
it has a serious vulnerability and is therefore not favorable.

A more secure approach is to restrict the conditions for removing an interval.
The interval can only be deleted after issuing a prepare transaction.
This prepare transaction triggers the resending
of all transactions from other entities.
These transactions are then re-included in the transaction memory pool.
After confirming the prepare transaction,
the interval can be deleted with the confirmation of the delete transaction.
The restricted authorized delete is the most complex method.

\subsection{Authorized Block Removal}
The restricted authorized block removal is illustrated in \cref{fig:chaindelexample}.
Here, Alice and Bob registered and send each a removable transaction (cf. State~1).
Their removable transactions are confirmed in the same interval, $I_{1}$.
Alice got a new e-mail address and wants to delete her old address stored in $sk_{A}(Rem(m))$.
However, Bob's transaction, $sk_{B}(Rem(n))$, should remain.
Therefore, Alice issues a prepare Transaction, $sk_{A}(Prep(1))$,
and a delete transaction, $sk_{A}(Del(1))$, for the interval.
This prepare transaction triggers the movement of all transactions
from other entities except Alice in this case $sk_{B}(Rem(n))$ from Bob, to the memory pool.
The confirmation of the prepare transaction confirms that
all necessary transactions are included in the previous interval.
After the prepare transaction is confirmed and
the removable transactions are therefore duplicated,
Alice's delete transaction for the interval can be confirmed in a block.
After safely confirming the delete transaction, the interval is deleted~(State 2).

The two-step removal process ensures faster verification times of
prepare and delete transactions.
The prepare transaction confirms the re-inclusion,
the condition for its validity.
The delete transaction is validated
by using the list of public keys in the interval.
If the list contains only the corresponding key(s) of the signer(s), it is valid.
Otherwise, the delete transaction needs the prepare transaction as an input.
Without the prepare transaction, the verification time
increases, requiring the check of every interval following the interval to be deleted.

While this approach prevents information loss by malicious deletes,
it reduces the throughput of removable transactions.
A malicious user who managed to insert removable transactions in many intervals
could issue many prepare transactions, effectively reissuing many personal transactions.
This might be mitigated by dynamically adjusting the interval length,
increasing the removable transactions throughput and/or by grouping
removable transactions to prevent the necessity of prepare transactions.
The simplest prevention of malicious deletes is to prevent
removable transactions of multiple entities in one interval.
Also, delete transactions can temporarily reduce the throughput in permanent blocks.
However, the amount of intervals is limited and
by temporarily reducing interval lengths to zero the chain can \enquote{catch up} with the deletes.

Removing intervals is not reversible.
Therefore, consensus algorithms without finality need
sufficient confirmations before deleting the interval.
After the successful delete, the second link from $B_{1}$
points to \enquote{nothing}.

In order to synchronize the blockchain, clients first retrieve permanent blocks
before filling the \enquote{gaps} with the still remaining removable blocks,
ensuring the verifiability as potential delete transactions can be processed first.

\section{Use Case: Consent Management}\label{sec:consent}
We propose a possible use case of our concept: consent management.
In our use case removable transactions serve to create a stronger connection
of the digital identity with the real-world identity of an entity.
While revealing further information about an entity is not strictly required
for the basic functionality of consent management,
it makes processing and collecting data easier.
Since in this case removable transactions contain personal data,
we require authorized deletes.
For the use case, we introduce two new transactions:
\emph{consent} and \emph{information transactions}.

For example, Alice, host of a website, would like to replace the HTTP-cookie banner.
Therefore, Alice issues an information transaction $sk_{A}(Info(Reg(A)))$.

$Info$ takes a register transaction's output as input,
contains information about the process,
and addresses the requirements for consent,
\ie, consent must be specific and informed.
Accordingly, it provides the required information
(\eg, data controller, recording purpose).
Moreover, purpose and categories are encoded in a binary encoding,
where each bit refers to a predefined purpose for which the data is used.
Therefore, $Info$ contains required information
as well as the encoding scheme, \eg,
bit one is for the strictly necessary cookies,
bit two is for the functional cookies and
bit three is for the performance cookies.

Bob wants to give his consent to Alice's cookies,
and sends a consent transaction.
The consent transaction takes Bob's register transaction as an input
and Alice's information transaction as an output, and has one open output.
The transferred value determines the agreement.
Now consider, Bob agrees only to the necessary cookies
and therefore sends $1$, \ie, $sk_{B}(Con_{1}(Reg(B),1))$.
After a while Bob also agrees to functional cookies and sends a new transaction.
The new consent transaction takes Bob's previous unspent consent transaction's output as an input,
Alice information transaction as an output and sends $3$, \ie, $sk_{B}(Con_{2}(Con_{1},3))$.
When Bob wants to revoke his consent, he sends a third consent transaction.
The third consent transaction takes the unspent output of the second consent transaction as an input,
Alice's information transaction as an output and sends a $0$, \ie, $sk_{B}(Con_{3}(Con_{2},0))$,
indicating the revoke of the consent.
After the output of Bob's first and second consent transaction are spent they loose their validity.
In general, consent transactions with an unspent transaction output (UTXO)
represent the current given consent.
By following the transaction graph,
the data controller can see who gave her the consent to use which information,
providing the required auditability.

At any time Alice and Bob can provide and remove additional information
as previously described.

\section{Discussion}\label{sec:analysis}
The removal method is determined by the use case,
but for personal data, authorized removal is preferable.
The use case can also specify which information should be stored in removable transactions,
but this is eventually decided by the transaction issuer.

For a P2P data storage and distribution protocol,
deletion in a distributed setting is difficult~\cite{politou2020delegated}.
Due to the existence of multiple copies, node churn, backups,
and a lack of control over other devices in general,
it is impossible to guarantee the deletion of all removable block copies.
In that aspect our protocol cannot solve the problem or guarantee GDPR compliance.
However, delete requests, \ie, delete transactions, are provably informing data controllers
to delete the data.

Furthermore, as similarly discussed by \citeauthor{puddu2017muchain}~\cite{puddu2017muchain}
transaction dependencies can raise consistency problems in mutable blockchains,
which we call \emph{deletion order}.
For a better understanding, why the deletion order has an impact,
let us assume we manage removable information with a state machine,
\eg, a smart contract.
A user interacts with the smart contract by issuing transactions
and therefore can change the state.
A sequence of transactions leads to a certain state.
Removing information, \ie, transactions,
can lead to a different state of the smart contract;
particularly when multiple entities are involved.
After a deletion, transactions of other entities are reissued,
resulting in a new transaction sequence,
which might yield a new state,
effectively violating consistency.

Since the deletion order is a fundamental problem,
we design removable transactions as independent transactions.
removing possible inconsistency issues.
Dependent transactions are possible, but require careful consideration.
The simplest approach would be to restrict
dependencies to one interval or enforcing a deletion order.

Our proposed chain structure has two different types of blocks.
The main difference in creating removable or permanent blocks
is the selection of included transaction types.
In a permissioned consortium setting this should not create any problems.
However, permissionless blockchains require additional considerations.
Transfer and mining of coins should only be possible in permanent blocks.
A miner would be required to mine interval and permanent block together.
This increases block creation and propagation time, reducing
willingness and success rate of mining longer intervals.
To increase willingness, mining intervals could have
decreased difficulty requirements or higher mining rewards.
In proof-of-work consensus algorithms,
removable blocks could require no additional work,
therefore, they do not include a nonce,
ensuring their release with the following permanent block.
 
Rewarding removable blocks should be carefully considered.
If the reward for removable blocks is high, a miner could try to create
a large interval and immediately delete the interval, simply for the reward.
By implementing a lock for deleting an interval, this problem can be mitigated.
A block mentioning a high interval length with no removable blocks can then be identified as faulty.
Thus, rewards for mining removable blocks should generally be low and
accounted in the following permanent block only.

Another aspect of our approach that needs to be considered,
is the introduced overhead, due to additional required data.
The overhead for the second link and the interval length is around $33\,B$.
To keep the header size constant in case of the authorized delete
a maximum amount of public keys can be defined, \eg, 4 keys.
Resulting in an overhead of $151\,B$, assuming $32\,B$ per key.

\section{Conclusion}\label{sec:conclusion}
In this work, we presented the concept of a
mutable blockchain and showed its utility for consent management.
In particular, the concept enables temporal storage of on-chain information,
while maintaining the possibility for erasure.
We discussed potential issues that stem from the structure and mutability.

\section*{Acknowledgments}
This work was supported by the Federal Ministry of Education and Research of Germany
 (project number 16KIS0909).

\IEEEtriggeratref{6}

\printbibliography[heading=bibintoc]

\end{document}